\documentclass[10pt]{article}

\usepackage{lmodern}
\usepackage[utf8]{inputenc}          % Acentos y caracteres especiales
\usepackage[T1]{fontenc}             % Codificación de fuentes
\usepackage{graphicx}                % Figuras
\usepackage{hyperref}                % Hipervínculos (DOI, URLs)
\usepackage{orcidlink}               % ORCID (opcional, requiere orcidlink.sty)
\usepackage{amsmath}

\begin{document}

\title{Shell-model description of collective deformation variables $\beta$ and $\gamma$ in $^{164-188}$Hf nuclei}

\author{Carlos E. Vargas\\
$^{1}$ Facultad de F\'isica, Universidad Veracruzana, Paseo No. 112, \\
Des. Hab. Nvo. Xalapa, Xalapa, 91097, Veracruz, Mexico \\
E-mail: cavargas@uv.mx}

\maketitle

\begin{abstract}

Symmetry-based models, such as the pseudo-SU(3) scheme, have proved to be highly useful tools for describing the low-energy structure of nuclei in the rare-earth region. This model is employed to study the evolution of the collective deformation variables $\beta$ and $\gamma$ in the ground state ($0^+$) along the isotopic chain $^{164-188}$Hf. The truncation is carried out by considering the irreducible representations (irreps) with high spatial symmetry and pseudo-spin 0 and 1, so that both the most and less symmetric relevant irreps are included, the latter being of particular importance for the description of low-energy properties, especially the shape transition. The Hamiltonian includes spherical Nilsson single-particle energies, the quadrupole–quadrupole interaction, and pairing correlations. It is parameterised in a systematic manner, so that the description is obtained without free parameters. In agreement with experimental data, the maximum deformation $\beta$ occurs at $N = 100$, although the model predicts values that are about 20\% smaller. As in other theoretical studies, a prolate deformation ($\gamma \sim 8^{\circ}$) is predicted for almost all nuclei in the shell, with a clear transition towards triaxial shapes for $N \geq 112$. An analysis of the effects of the pairing and Nilsson terms on the values of $\beta$ and $\gamma$ indicates that it is the latter that determine the prolate–oblate transition, with the pairing term playing only a marginal role. The absence of nucleons in the intruder sector is the most significant limitation of the model.

\end{abstract}

\section{Introduction}\label{sec-intro}

The rotational degrees of freedom of the nucleus are associated with a deformed equilibrium state, which is a consequence of the shell structure. Many nuclei exhibit a permanent deformation, often of quadrupole character. Such quadrupole shapes are described in terms of the parameters $\beta$ and $\gamma$: the first defines the magnitude of the deformation, while $\gamma$ specifies the degree of axial asymmetry or triaxiality \cite{Boh98}.

Both experimental data and theoretical results indicate that most deformed nuclei exhibit a prolate quadrupole deformation. However, some regions of the nuclear chart display oblate shapes or even deformed mass distributions that break axial symmetry \cite{Rob09,Law25}. 

Triaxial nuclear deformation is relevant to the formation of strongly deformed bands \cite{Ode01,Var25} associated with wobbling-type excitations, to the creation of heavy elements \cite{Gro14} via neutron capture in stellar explosions, and to the determination of low-energy nuclear structure, where some nuclei even exhibit multiple shape coexistence \cite{Gar19}, among other phenomena. Such deformation is inferred mainly through indirect methods that compare model predictions with experimental measurements. Examples include: (a) the fragmentation of relativistic projectiles \cite{Caa05}, (b) the isovector giant dipole resonance \cite{Men21}, (c) signature splitting and inversion in rotational bands \cite{Ben84}, and (d) tilted-axis rotation in triaxial nuclei \cite{Fra97}.

From the theoretical point of view, the prolate–oblate transition in heavy rare-earth nuclei is a well-established phenomenon. Among several studies we can mention those based on phenomenological algebraic models, such as the IBM \cite{Bon24,Jol03}, the proxy-SU(3) scheme \cite{Bon17,Bon23}, the mapping method that derives the IBM Hamiltonian from microscopic HFB calculations \cite{Nom11}, and, within a mean-field framework, Skyrme-HF \cite{Ste05} and Gogny-HFB \cite{Rob09} calculations. These models consistently indicate that this is a first-order nuclear phase transition that occurs around $N = 116$, characterised by pronounced $\gamma$-softness rather than rigid triaxiality. 

In order to provide elements that contribute to the understanding of this transition, we employ the pseudo-SU(3) coupling scheme, which allows us to describe the collective deformation variables along the isotope chain $^{164-188}$Hf. This isotopic chain was chosen because it lies in the rare-earth region, where nuclei are well deformed and suitable for a pseudo-SU(3) description. Importantly, this chain is expected to undergo a prolate-to-oblate shape transition as a function of neutron number, passing through triaxial shapes, offering a unique testing ground for our extended framework.

Originally developed by Elliott in 1958, the SU(3) model \cite{Ell58,Ell58b} proved to be very useful for studying the structure of light nuclei. However, in heavy nuclei the strong spin–orbit interaction breaks the harmonic-oscillator symmetry. At the same time, the pseudo-spin symmetry emerges \cite{Ari69,Hec69}. This refers to the experimentally observed fact, of relativistic origin \cite{Gin97}, that single-particle orbitals with $j=l-1/2$ and $j=(l-2)+1/2$ in the shell $\eta$ lie very close in energy. This proximity allows them to be classified as pseudo-spin doublets with quantum numbers $\tilde{j}=j$, $\tilde{\eta}=\eta-1$, and $\tilde{l}=l-1$, giving rise to the pseudo-SU(3) model \cite{Rat73}.

The first calculations \cite{Dra84,Cas87} employed the model as a dynamical symmetry, using the leading irreducible representation (irrep) to describe the properties of the ground-state band. This modelling strategy has also been used to predict the collective deformation variables $\beta$ and $\gamma$ in many rare-earth nuclei, from $^{138}$Xe to $^{202}$Pt \cite{Bon20}. In these studies, the description was carried out by considering only the irrep with maximum spatial symmetry for each nucleus and, in some cases, also the next irrep with the highest spatial symmetry.

The development of sophisticated computer codes to calculate triply reduced matrix elements \cite{Bah94} has made it possible to analyse the interplay between the quadrupole–quadrupole and spin–orbit interactions \cite{Var98}, and to study the combined effects of the quadrupole–quadrupole, pairing, and spin–orbit terms on ground-state shapes \cite{Esc98}. In addition, it has enabled calculations with realistic Hamiltonians, including terms that break pseudo-spin symmetry, parametrised through a normalisation mechanism governed by a universal $A^{-1/3}$ scaling \cite{Duf96,Rin79}. This has allowed the use of the pseudo-SU(3) shell model in heavy nuclei, in which the symmetry-breaking terms mix the irreps and smooth out the abrupt variations predicted by the dynamical symmetry along the shell \cite{Var00,Var02,Var04,Var13,Var17,Var24,Var25}.

In this contribution, we present the pseudo-SU(3) model results for the geometric variables $\beta$ and $\gamma$ in the nuclei $^{164-188}$Hf. The Hilbert space is constructed through a truncation based on the symmetry of the states, and a realistic Hamiltonian is employed for the interaction, including the quadrupole–quadrupole term as well as the Nilsson and pairing symmetry-breaking terms. A detailed analysis is also presented of the effects produced by the Nilsson single-particle and pairing terms on the values of the geometric variables $\beta$ and $\gamma$ along the isotope chain under consideration.

In Sec. \ref{sec-base} the construction of the basis states for a many-body system is discussed, and an example is given in Sec. \ref{sec-wf} for the nucleus $^{174}$Hf. In Sec. \ref{sec-ham} the Hamiltonian is presented and the ground-state wave function is analysed, again taking $^{174}$Hf as an example. In Sec. \ref{sec-bega}, the main results are presented, namely the $\beta$ and $\gamma$ values of the $0^+$ ground state along the $^{164-188}$Hf isotope chain. In Sec. \ref{sec-pair} the relevance of the Nilsson energies and pairing terms for the deformation is discussed. Finally, the conclusions are summarised in Sec. \ref{sec-conclu}.

\section{Construction of the Model Space}\label{sec-base}

 The starting point of the theoretical description within the pseudo-SU(3) model is the construction of the many-body basis from the most probable distribution of protons and neutrons in the normal- and intruder-parity shells. Using a Nilsson level diagram \cite{Dra84,Cas87}, we determine the valence protons and neutrons for a given deformation of the nucleus under study. To obtain the ground-state nuclear deformations, we follow the estimates of the finite-range droplet macroscopic model combined with the folded-Yukawa single-particle microscopic model for 8979 nuclei across the whole chart of nuclides \cite{Mol95}, in the same way as in previous applications of the model \cite{Var00}. For the Hafnium isotopes, there are 72 protons, 22 of which occupy the $\tilde{\eta}=3$ shell; of these, 14 are distributed over normal-parity orbitals and 8 over intruder orbitals. For the neutrons, the occupation numbers assigned to each nucleus are given in Table~\ref{occup}.
\begin{table}[h]
\begin{center} \begin{tabular}{c|c|cc}
Nuclei & $\epsilon_2$ & $n_\nu^N$ & $n_\nu^A$
\\ \hline
$^{164}$Hf$_{92}$ & 0.192 & 8  &  2 \\
$^{166}$Hf$_{94}$ & 0.208 & 8  &  4 \\
$^{168}$Hf$_{96}$ & 0.233 & 10 &  4 \\
$^{170}$Hf$_{98}$ & 0.250 & 10 &  6 \\
$^{172}$Hf$_{100}$& 0.258 & 12 &  6 \\
$^{174}$Hf$_{102}$& 0.258 & 14 &  6 \\
$^{176}$Hf$_{104}$& 0.250 & 14 &  8 \\
$^{178}$Hf$_{106}$& 0.250 & 16 &  8 \\
$^{180}$Hf$_{108}$& 0.250 & 16 & 10 \\
$^{182}$Hf$_{110}$& 0.242 & 18 & 10 \\
$^{184}$Hf$_{112}$& 0.233 & 20 & 10 \\
$^{186}$Hf$_{114}$& 0.225 & 22 & 10 \\
$^{188}$Hf$_{116}$& 0.192 & 24 & 10 \\
\end{tabular}\end{center}
\caption{Deformation ($\epsilon_2$) \cite{Mol95} and occupation numbers for neutrons. $N$ and $A$ indicate normal and abnormal parity levels, respectively.}
\label{occup}\end{table}

It is important to mention that, as in previous works, the nucleons in intruder-parity orbitals renormalise the dynamics described by the nucleons in normal-parity levels, which has required the use of effective charges larger than those usually employed in shell-model calculations. Although this has been a reasonable approximation, it is a very strong assumption and represents the most important limitation of the present version of the theory.

In previous works, the construction of many-body states for \(n_\alpha\) active nucleons in a given normal-parity shell has been presented in detail \cite{Var04,Var25}. The rule for determining the allowed K and L values for a given $(\lambda,\mu)$ irrep can be found in Harvey's 1968 review on the SU(3) model \cite{Har68}. A quantum state \(|J_i M\rangle\), where \(J\) is the total angular momentum, \(M\) its projection, and \(i\) an integer that labels the states with the same \(J,M\), is constructed as a linear combination
\begin{equation}
| J_i M \rangle = \sum_\beta C^{JM}_\beta |\beta J M \rangle \label{eq:wf}
\end{equation}

\noindent of coupled proton–neutron states

\begin{eqnarray}
|\beta JM \rangle \equiv \nonumber \\
| \{ \tilde{f}_\pi \} (\lambda_\pi \mu_\pi) \tilde{S}_\pi, \{ \tilde{f}_\nu \}
(\lambda_\nu \mu_\nu) \tilde{S}_\nu ; \rho (\lambda \mu ) \kappa L, 
\tilde{S}~ JM \rangle  =
\nonumber \\
   \sum_{M_L M_S} (L M_L, \tilde{S} M_S | J M )
\sum_{M_{S \pi} M_{S \nu}}
(\tilde{S}_\pi M_{S \pi}, \tilde{S}_\nu M_{S \nu} | \tilde{S} M_S) 
\nonumber \\
\sum_{ \begin{array}{c} k_\pi L_\pi M_\pi \\ \kappa_\nu L_\nu M_\nu 
\end{array} }
{ \langle (\lambda_\pi \mu_\pi) \kappa_\pi L_\pi M_\pi ;
(\lambda_\nu \mu_\nu) \kappa_\nu L_\nu M_\nu |
(\lambda \mu ) \kappa L M \rangle}_\rho \nonumber\\
| \{ \tilde{f}_\pi \} (\lambda_\pi \mu_\pi) \kappa_\pi L_\pi M_\pi, 
\tilde{S}_\pi M_{S
\pi}  \rangle  ~~~~~~~~~~~~~~\label{basis}  \\
| \{ \tilde{f}_\nu \} (\lambda_\nu \mu_\nu) \kappa_\nu L_\nu M_\nu, 
\tilde{S}_\nu M_{S
\nu} \rangle  ~~~~~~~~~~~~~~~.\nonumber
\end{eqnarray}

\noindent In the above expression, $\langle-;-|- \rangle$ and $(-,-|-)$ are the SU(3) and SU(2) Clebsch Gordan coefficients, respectively.

As mentioned before, the use of the model as a dynamical symmetry \cite{Dra84,Cas87}, considering only the dominant irrep with pseudo-spin 0 for an even number of protons or neutrons, has made it possible to study the collective deformation parameters $\beta$ and $\gamma$ over wide regions of the nuclear chart \cite{Bon20}. In those studies, only the most stretched irrep $(\lambda,\mu) = (\lambda_\pi+\lambda_\nu,\mu_\pi+\mu_\nu)$ has been considered, where $(\lambda_\pi,\mu_\pi)$ and $(\lambda_\nu,\mu_\nu)$ are the proton and neutron irreps, respectively, corresponding to the states with the highest spatial symmetry. However, a more realistic description that includes a symmetry-breaking Hamiltonian requires a much larger Hilbert space, which must be truncated.

The truncation scheme used in recent applications of this model \cite{Var24,Var25} consists of ordering the SU(3) irreps, starting from those with the largest eigenvalue of the second‑order Casimir operator \(C_2\). This procedure is applied to protons and neutrons separately. The irreps thus obtained are then coupled in order to generate a set of nuclear states, which are also ordered according to their \(C_2\) eigenvalue. This truncation scheme has worked satisfactorily for nuclei in which both protons and neutrons occupy orbitals below mid-shell, since it selects the more deformed and symmetric irreps, which are suitable for describing low-energy properties. However, the scheme breaks down for nuclei with a number of protons and/or neutrons beyond mid-shell. The short-range nature of the nucleon–nucleon interaction forces the spatial part of the nuclear wave function to be as symmetric as possible \cite{Isa95}, within the limits imposed by the Pauli exclusion principle. Consequently, instead of selecting the maximally deformed states, it becomes essential to use a truncation criterion based on the highest spatial symmetry of the proton and neutron irreps, as well as of their coupled configurations.

It is necessary to make a clarification regarding the nomenclature used to label the states with maximum spatial symmetry (or “highest-weight states”, as they are called in Ref.~\cite{Bon17}). 

In the SU(3) scheme developed by J. P. Elliott in 1958 \cite{Ell58,Ell58b}, the highest-weight state is defined algebraically as the one that is annihilated by the raising operators of the SU(3) group in the decomposition $SU(3) \supset SU(2) \times U(1)$, being unique and fundamental for constructing the basis. The usefulness of this classification lies in avoiding repetition of equivalent rotated states by choosing the highest-weight state within the representation. However, the term “highest-weight states (hws)" has been used \cite{Bon17} to refer to the states with maximum orbital symmetry, characterised by maximum values of $(\lambda+\mu)$ and of the length of the first row of the Young diagram, and to select between irreps with different symmetries. 

The distinction can be illustrated with a single particle in the $2\hbar \omega$ shell. Under U(3), the three configurations (2,0,0), (0,2,0) and (0,0,2) represent the same intrinsic state, rotated in space. Elliott's decomposition $SU(3) \supset SU(2) \times U(1)$ selects the one with the largest number of quanta in the z-direction, and then in the x-direction, as the unique highest-weight state: (2,0,0), corresponding to ($\lambda,\mu$)=(2,0). The other two are merely rotated copies. In contrast, (2,0,0) and (1,1,0) are not rotationally equivalent; they have different spatial symmetries, with SU(3) irreps (2,0) and (0,1). Here, (2,0) has maximum spatial symmetry (larger $\lambda+\mu$) and is selected on physical grounds. However, calling (2,0) a highest-weight state relative to (0,1) would be incorrect: the algebraic definition applies strictly within rotationally equivalent states, whereas spatial symmetry compares states from distinct symmetry classes.

The usage of this terminology in Ref. \cite{Bon17} deviates from the standard group-theoretic definition. To avoid confusion with the standard terminology in group theory and to emphasise its physical meaning, in this work we systematically refer to the “highest-weight states” mentioned in Ref.~\cite{Bon17} as states with maximum spatial symmetry. 

Because of the points discussed above, the study of the structure of rare-earth nuclei requires the consideration of states with high spatial symmetry, so in this work the truncation has been carried out following the prescription presented by D. Bonatsos et al. \cite{Mar21}. However, this prescription has been formulated by treating the model as a dynamical symmetry, taking only the most symmetric state and, as an extension of this, the second (nhw) or third (nnhw) that follow it \cite{Bon24}. A more general study involves considering a much larger number of irreps with a realistic Hamiltonian that breaks the symmetry, in which the ground state is formed as a linear combination of SU(3) eigenstates with high spatial symmetry. 

In the next Section, we discuss the criteria used to select the more symmetric irreps. 

\section{Wave function for $^{174}$Hf}\label{sec-wf}

The spatial symmetry of the wave function plays a fundamental role, because the short‑range character of the nucleon–nucleon interaction favours the more symmetric irreducible representations of the pseudo‑SU(3) model. Whereas a previous study \cite{Bon20} treats pseudo‑SU(3) as a dynamical symmetry, taking only the irrep with maximum spatial symmetry, in the present work we adopt a more general approach. We start from a truncation of the Hilbert space that selects the more symmetric SU(3) irreps and then we diagonalise a realistic Hamiltonian without free parameters in the selected space. The resulting states, which are mixtures of different irreps, allow us to calculate the collective observables $\beta$ and $\gamma$, providing a more realistic description of the nuclear structure that explicitly incorporates the mixing of irreps induced by the symmetry‑breaking terms.

To carry out the truncation of the Hilbert space, we use the two‑fold condition \cite{Mar21}, which allows us to determine the irreps with the highest spatial symmetry and the largest fraction of symmetric quanta. The criterion is based on the value of $\lambda + \mu$ (the number of boxes in the first row of the Young diagram) and on \begin{equation}
r(\lambda,\mu)=\frac{\lambda+\mu}{\lambda+2\mu}\cdot 100\%~,
\label{r}\end{equation}
which measures the degree of spatial symmetry.

In order to illustrate how the basis is built, we take $^{174}$Hf as an example. This nucleus has 22 valence protons; of these, 14 occupy normal-parity orbitals in the $\tilde{\eta}=3$ shell, while the remaining 8 lie in the intruder sector. Out of the 48 possible irreps with $\tilde{S}_\pi = 0,1$, Table~\ref{p-irr}
\begin{table}[h]
\begin{center} 
\begin{tabular}{cccccc||cccccc}
$\tilde{S}_\pi$ & $\lambda$ & $\mu$ & $\lambda + \mu$ & r$(\%)$ & $C_2$ & $\tilde{S}_\pi$ & $\lambda$ & $\mu$ & $\lambda + \mu$ & r$(\%)$ & $C_2$ \\ \hline
0 &\bf6&\bf6& 12 &\bf66&144 & 1 & 3 & 9  &  12 & 57 & 153 \\
  & 3 & 9  &  12 & 57 & 153 &   &\bf7&\bf4& 11 &\bf73&126 \\
  & 0 & 12 &  12 & 50 & 180 &   & 4 & 7  &  11 & 61 & 126 \\
  & 4 & 7  &  11 & 61 & 126 &   & 1 & 10 &  11 & 52 & 144 \\
  &\bf8&\bf2& 10 &\bf83&114 &   & 5 & 5  &  10 & 66 & 105 \\
  & 5 & 5  &  10 & 66 & 105 &   & 2 & 8  &  10 & 55 & 114 \\
  & 2 & 8  &  10 & 55 & 114 &   &\bf9&\bf0&  9&\bf100&108 \\
  & 6 & 3  &   9 & 75 &  90 &   & 6 & 3  &   9 & 75 &  90 \\
  & 3 & 6  &   9 & 60 &  90 &   & 3 & 6  &   9 & 60 &  90 \\
  &   &    &     &    &     &   & 0 & 9  &   9 & 50 & 108 \\
\end{tabular}\end{center}
\caption{Shown are the 19 most symmetric $\tilde{SU}(3)$ irreps for 14 protons in the $\tilde{\eta}=3$ shell. The meaning of the columns is as follows. The first gives the pseudo‑spin 0; the second and third give the irreps $\lambda$ and $\mu$, respectively. The fourth shows the value of $\lambda + \mu$; the fifth gives the value of the ratio $r$ (Eq. \ref{r}); and the sixth shows the value of $C_2$. From the seventh to the twelfth column, the same quantities are given for irreps with pseudo‑spin 1. In bold, we highlight the irreps with the largest percentage of symmetric quanta.}
\label{p-irr}\end{table}
shows the more symmetric ones, ordered according to their pseudo-spin ($\tilde{S}_\pi = 0,1$), decreasing values of $\lambda + \mu$, the value of $r(\%)$, and, for reference, the corresponding $C_2$ eigenvalue. 

The first three irreps shown in Table \ref{p-irr} with $\tilde{S}_\pi = 0$ and $(\lambda + \mu)_{max} = 12$ — (6,6), (3,9) and (0,12) — correspond exactly to the three more symmetric irreps listed in Table 3 of Ref. \cite{Mar21} for 14 particles with U(10) symmetry (referred to as the highest weight (hw), the next-to-highest weight (nhw), and (n$^2$hw), respectively). However, in order to build a broader basis of states that can mix when the symmetry is broken, it is necessary to consider other irreps with $\tilde{S}_\pi = 1$ and $\lambda' + \mu' < (\lambda + \mu)_{max}$. In Table \ref{p-irr}, the irreps with the largest percentage $r$ of symmetric quanta are highlighted in bold, since they play a major role in the description of the ground state.

To continue with the nucleus $^{174}$Hf as an example, we now consider the neutrons. Out of the 102 neutrons in this nucleus, 20 occupy the $\tilde{\eta}=4$ shell, of which 14 are distributed over normal-parity orbitals and 6 over intruder orbitals. In Table \ref{n-irr}
\begin{table}[h]
\begin{center} \begin{tabular}{cccccc||cccccc}
$\tilde{S}_\nu$ & $\lambda$ & $\mu$ & $\lambda + \mu$ & r$(\%)$ & $C_2$ &  $\tilde{S}_\nu$ & $\lambda$ & $\mu$ & $\lambda + \mu$ & r$(\%)$ & $C_2$ \\ \hline
0 &\bf20&\bf6& 26 &\bf81&634 & 1 & 17 &  9 &  26 & 78 & 601 \\
  & 17 & 9  &  26 & 74 & 601 &   & 14 & 12 &  26 & 68 & 586 \\
  & 14 & 12 &  26 & 68 & 586 &   & 11 & 15 &  26 & 63 & 589 \\
  & 11 & 15 &  26 & 63 & 589 &   &\bf21&\bf4& 25&\bf86& 616 \\
  & 8  & 18 &  26 & 59 & 610 &   & 18 &  7 &  25 & 78 & 574 \\
  & 18 & 7  &  25 & 78 & 574 &   & 15 & 10 &  25 & 71 & 550 \\
  & 15 & 10 &  25 & 71 & 550 &   & 12 & 13 &  25 & 65 & 544 \\
  & 12 & 13 &  25 & 65 & 544 &   &  9 & 16 &  25 & 60 & 556 \\
  &  9 & 16 &  25 & 60 & 556 &   &  6 & 19 &  25 & 56 & 586 \\
  &\bf22&\bf2& 24&\bf92& 604 &   & 19 &  5 &  24 & 82 & 553 \\
  & 19 &  5 &  24 & 82 & 553 &   & 16 &  8 &  24 & 75 & 520 \\
  & 16 &  8 &  24 & 75 & 520 &   & 13 & 11 &  24 & 68 & 505 \\
  & 13 & 11 &  24 & 68 & 505 &   & 10 & 14 &  24 & 63 & 508 \\
  & 10 & 14 &  24 & 63 & 508 &   &  7 & 17 &  24 & 58 & 529 \\
  &  7 & 17 &  24 & 58 & 529 &   &  4 & 20 &  24 & 54 & 568 \\
  &  4 & 20 &  24 & 54 & 568 &   &\bf23&\bf0& 23&\bf100&598 \\
  & 20 &  3 &  23 & 88 & 538 &   & 20 &  3 &  23 & 88 & 538 \\
  & 17 &  6 &  23 & 79 & 496 &   & 17 &  6 &  23 & 79 & 496 \\
  & 14 &  9 &  23 & 71 & 472 &   & 14 &  9 &  23 & 71 & 472 \\
  & 11 & 12 &  23 & 65 & 466 &   & 11 & 12 &  23 & 65 & 466 \\
  &  8 & 15 &  23 & 60 & 478 &   &  8 & 15 &  23 & 60 & 478 \\
  &  5 & 18 &  23 & 56 & 508 &   &  5 & 18 &  23 & 56 & 508 \\
  &    &    &     &    &     &   &  2 & 21 &  23 & 52 & 556 \\
\end{tabular}\end{center}
\caption{Displayed are the 45 more symmetric $\tilde{SU}(3)$ irreps for 14 neutrons in the $\tilde{\eta}=4$ shell. The column definitions follow the same convention as in Table \ref{p-irr}. Irreps with the largest percentage ($r$) of symmetric quanta are indicated in boldface.}
\label{n-irr}\end{table}
we present the more symmetric irreps, ordered according to their $\tilde{S}_\nu$, the value of $\lambda+\mu$, the ratio $r$, and the value of $C_2$. The 45 more symmetric irreps are taken from the 228 that make up the full neutron space with $\tilde{S}_\nu = 0$ and 1. 

The first five rows on the left-hand side show the most symmetric irreps with $\tilde{S}_\nu = 0$ and $(\lambda+\mu)_{\text{max}} = 26$, which coincide with those listed in Table 4 of Ref.~\cite{Mar21} for 14 particles with U(15) symmetry.

Taking the proton and neutron irreps listed in Tables \ref{p-irr} and \ref{n-irr}, respectively, we generate the basis for $^{174}$Hf. Table \ref{irr}
\begin{table}[h]
\begin{center} \begin{tabular}{ccc|cccc}
$\lambda$ & $\mu$ & $\tilde{S}$ & $\lambda + \mu$ & $r$(\%) & $C_2$ & \#(mult.)\\ \hline
26 & 12 &   0 & 38 & 76 & 1246 &  1 \\
23 & 15 & 0,1 & 38 & 71 & 1213 &  4 \\
20 & 18 &0,1,2& 38 & 67 & 1198 &  9 \\
17 & 21 &0,1,2& 38 & 64 & 1201 & 10 \\
14 & 24 &0,1,2& 38 & 61 & 1222 &  9 \\
27 & 10 & 0,1 & 37 & 78 & 1210 &  3 \\
24 & 13 &0,1,2& 37 & 74 & 1168 & 17 \\
21 & 16 &0,1,2& 37 & 69 & 1114 & 33 \\
28 &  8 &0,1,2& 36 & 81 & 1180 &  8 \\
25 & 11 &0,1,2& 36 & 76 & 1129 & 35 \\
29 &  6 &0,1,2& 35 & 85 & 1156 & 12 \\
30 &  4 &0,1,2& 34 & 89 & 1138 & 17 \\
31 &  2 &0,1,2& 33 & 94 & 1126 & 11 \\
32 &  0 &0,1,2& 32 &100 & 1120 &  8 \\
\end{tabular}\end{center}
\caption{ The 177 irreps for $^{174}$Hf are presented. The first three columns give the nuclear irrep labels $\lambda$ and $\mu$, and the corresponding pseudo-spin. The fourth to sixth columns show the value of $\lambda + \mu$, the ratio $r(\lambda,\mu)$ (Eq.~\ref{r}), and $C_2$, respectively. The seventh column gives the multiplicity associated with each irrep $(\lambda,\mu)$.}
\label{irr}\end{table}
shows the 177 more symmetric irreps for the nucleus, obtained by taking the direct product of the proton and neutron irreps. 

We must stress the great importance of including pseudo-spin 1 states for the proton and/or neutron irreps \cite{Zuk95}. In constructing the final wave function, we find that practically all nuclear irreps $(\lambda,\mu)\tilde{S}$ (except for the most stretched irrep $(26,12)0=(6,6)0 \otimes (20,6)0$) are formed with irreps having $\tilde{S}_\alpha = 1$ ($\alpha=\pi$ or $\nu$), so that many nuclear irreps have a non-zero total pseudo-spin. In earlier works with the model \cite{Var04}, the relevance of states with $\tilde{S}=1$ for a proper description of the low-energy properties of rare-earth nuclei has been discussed, which has in some cases allowed the description of up to nine excited bands of the $^{160}$Dy nucleus.

\section{The Hamiltonian}\label{sec-ham}

The realistic collective interaction used includes the spherical Nilsson single-particle terms for protons ($H_{sp,\pi}$) and neutrons ($H_{sp,\nu}$), the quadrupole–quadrupole interaction ($\tilde{Q} \cdot \tilde{Q}$), and the pairing interactions ($H_{pair,\pi}$ and $H_{pair,\nu}$), all parametrised systematically,

\begin{eqnarray}
H_0 & = & \sum_{\alpha=\pi,\nu} \{ H_{sp,\alpha} - G_\alpha 
~H_{pair,\alpha} \} - \frac{1}{2}~  \chi~ \tilde Q \cdot \tilde Q ~.
\label{eq:ham}
\end{eqnarray}

In their seminal 1996 work, Dufour and Zuker \cite{Duf96} derived an effective collective Hamiltonian from realistic nucleon–nucleon forces. They demonstrated that, once the monopole field is removed, the residual multipole interaction is overwhelmingly dominated by separable pairing and quadrupole terms, while higher‑order multipoles can be neglected to a good approximation. Moreover, they showed that the normalisations of these leading terms are fixed by a universal $A^{-1/3}$ scaling, thereby eliminating any arbitrary dependence of the coupling strengths on the model space and establishing a realistic shell‑model Hamiltonian with fundamentally defined parameters, suitable for the systematic description of nuclear spectroscopy. Such a Hamiltonian has been employed in previous work within the model, using the aforementioned systematic parametrisation \cite{Var-00}, showing that it is possible to construct a parameter-free model for describing the structure of heavy nuclei.

The single-particle terms ($H_{sp,\alpha}$) can be expressed by
\begin{eqnarray}
H_{sp,\alpha} = \sum_{i_\alpha} \left( \hbar \omega_0 (\hat{\eta}_{i_\alpha}+\frac{3}{2})
+ C_\alpha {\bf  l}_{i_\alpha} \cdot {\bf s}_{i_\alpha} + D_\alpha {\bf  l}^2_{i_\alpha}\right),
~~~~~\alpha=\pi,\nu,
\label{eq-hsp}
\end{eqnarray}
where $C_\alpha$ and $D_\alpha$ are determined following the usual prescriptions \cite{Rin79}. In the pseudospin basis the spin–orbit and orbit–orbit contributions are small, yet they are responsible for most of the mixing between pseudo-SU(3) irreps. The pairing interaction also breaks the symmetry, but it plays a comparatively minor role in generating this mixing.

Once the basis and the Hamiltonian have been chosen, we can calculate the mixing of irreps that makes up the $J^\pi = 0^+$ ground state and analyse its wave function. Taking again $^{174}$Hf as an example, Table \ref{wf}
\begin{table}[h]
\begin{center} \begin{tabular}{ccc||ccc|ccc||c}
\multicolumn{3}{c||}{Total}&\multicolumn{3}{c|}{Protons}&\multicolumn{3}{c||}{Neutrons}& \\
$\lambda$ & $\mu$ & $\tilde{S}$ & $\lambda$ & $\mu$ & $\tilde{S}_\pi$ & $\lambda$ & $\mu$ & $\tilde{S}_\nu$ & \% \\ \hline
26 & 12 & 0 & 6 & 6 & 0 & 20 & 6 & 0 & 1.22 \\
27 & 10 & 1 & 6 & 6 & 0 & 21 & 4 & 1 & 0.88 \\
27 & 10 & 1 & 7 & 4 & 1 & 20 & 6 & 0 & 2.31 \\
28 &  8 & 0 & 6 & 6 & 0 & 20 & 6 & 0 & 0.42 \\
28 &  8 & 0 & 6 & 6 & 0 & 22 & 2 & 0 & 3.24 \\
28 &  8 & 0 & 7 & 4 & 1 & 21 & 4 & 1 & 0.62 \\
28 &  8 & 2 & 7 & 4 & 1 & 21 & 4 & 1 & 1.10 \\
28 &  8 & 0 & 8 & 2 & 0 & 20 & 6 & 0 & 6.85 \\
29 &  6 & 1 & 6 & 6 & 0 & 21 & 4 & 1 & 0.35 \\
29 &  6 & 1 & 6 & 6 & 0 & 23 & 0 & 1 & 1.22 \\
29 &  6 & 1 & 7 & 4 & 1 & 20 & 6 & 0 & 0.89 \\
29 &  6 & 1 & 7 & 4 & 1 & 22 & 2 & 0 & 6.44 \\
29 &  6 & 1 & 8 & 2 & 0 & 21 & 4 & 1 & 5.22 \\
29 &  6 & 1 & 9 & 0 & 1 & 20 & 6 & 0 & 3.39 \\
30 &  4 & 0 & 6 & 6 & 0 & 20 & 6 & 0 & 0.24 \\
30 &  4 & 0 & 6 & 6 & 0 & 22 & 2 & 0 & 1.87 \\
30 &  4 & 0 & 7 & 4 & 1 & 21 & 4 & 1 & 0.26 \\
30 &  4 & 2 & 7 & 4 & 1 & 21 & 4 & 1 & 0.47 \\
30 &  4 & 0 & 7 & 4 & 1 & 23 & 0 & 1 & 0.87 \\
30 &  4 & 2 & 7 & 4 & 1 & 23 & 0 & 1 & 1.53 \\
30 &  4 & 0 & 8 & 2 & 0 & 20 & 6 & 0 & 3.89 \\
30 &  4 & 0 & 8 & 2 & 0 & 22 & 2 & 0 & 19.48\\
30 &  4 & 0 & 9 & 0 & 1 & 21 & 4 & 1 & 0.94 \\
30 &  4 & 2 & 9 & 0 & 1 & 21 & 4 & 1 & 1.66 \\
31 &  2 & 1 & 6 & 6 & 0 & 21 & 4 & 1 & 0.28 \\
31 &  2 & 1 & 7 & 4 & 1 & 20 & 6 & 0 & 0.65 \\
31 &  2 & 1 & 7 & 4 & 1 & 22 & 2 & 0 & 3.54 \\
31 &  2 & 1 & 8 & 2 & 0 & 21 & 4 & 1 & 2.89 \\
31 &  2 & 1 & 8 & 2 & 0 & 23 & 0 & 1 & 6.72 \\
31 &  2 & 1 & 9 & 0 & 1 & 22 & 2 & 0 & 9.00 \\
32 &  0 & 0 & 6 & 6 & 0 & 20 & 6 & 0 & 0.16 \\
32 &  0 & 0 & 7 & 4 & 1 & 21 & 4 & 1 & 0.17 \\
32 &  0 & 2 & 7 & 4 & 1 & 21 & 4 & 1 & 0.29 \\
32 &  0 & 0 & 8 & 2 & 0 & 22 & 2 & 0 & 8.69 \\
32 &  0 & 0 & 9 & 0 & 1 & 23 & 0 & 1 & 0.78 \\
32 &  0 & 2 & 9 & 0 & 1 & 23 & 0 & 1 & 1.38 \\
\end{tabular}\end{center}
\caption{Pseudo-SU(3) components of the calculated ground state. The first three columns give the $(\lambda,\mu)\tilde{S}$ labels for $^{174}$Hf. Columns four to six list the proton irreps and pseudo-spin, columns seven to nine list the neutron irreps and pseudo-spin, and the tenth column gives the percentage with which each irrep contributes to the ground state.}
\label{wf}\end{table}
lists the 36 more symmetric pseudo-SU(3) irreps and their contributions to the ground state.

The percentage contributed by each irrep to the $0_1^+$ state is calculated from the wave function, Eq.~(\ref{eq:wf}), simply as $|C_\beta^{JM}|^2 \times 100~\%$. Given the large number of irreps in the truncated basis, we show only those whose contributions are larger than 0.1\%, which together account for 99.91\% of the total wave function.

\section{Results for shape variables}\label{sec-bega}

Earlier studies within the pseudo-SU(3) scheme \cite{Cas88,Dra89,Var25} have shown the relation between the model labels $\lambda$ and $\mu$ and the shape variables $\beta$ and $\gamma$ of the Bohr–Mottelson model \cite{Boh98}. This relation is obtained by mapping the invariants of the two theories onto each other. The resulting correlation allows us to assign unique values of the geometric variables $\beta$ and $\gamma$ to each SU(3) irrep $(\lambda,\mu)$ through the relations:

\begin{eqnarray}
\begin{aligned}
& \beta^2 =\frac{4\pi}{5} \frac{1}{(Ar_{rms}^2)^2}(\lambda^2 + 
  \lambda \mu + \mu^2 + 3\lambda + 3\mu + 3),\\ 
& \gamma = \mathrm{arctan} \left[ \frac{\sqrt{3}(\mu + 1)}{2 
  \lambda + \mu + 3} \right].
\label{eq:beta-gamma}
\end{aligned}
\end{eqnarray}

The interested reader is referred to the cited works for further details.

Since we only consider valence nucleons, the value of $\beta$ must be rescaled with respect to the size of the shell \cite{Esc98,Bon20}. In the present case, $\beta$ must be multiplied by the factor $A/[2(\mathcal{N}_\pi+\mathcal{N}_\nu)]$, where $\mathcal{N}_\pi=10$ and $\mathcal{N}_\nu=15$ are the numbers of spatial orbitals in the proton and neutron valence shells, respectively, so that the rescaling factor for $\beta$ is $A/50$. 

\begin{figure}[h]
\centering
\includegraphics[width=0.66\textwidth]{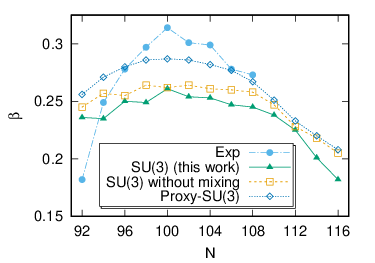}
\caption{Pseudo-SU(3) model predictions (labelled as “SU(3) (this work)”) for the collective deformation variable $\beta$ along the $^{164-188}$Hf isotopic chain, compared with the experimental data (“Exp”) \cite{Pri16}, the results obtained considering only the most symmetric pseudo-SU(3) representation (“SU(3) without mixing”) \cite{Bon20}, and the results from the proxy-SU(3) scheme \cite{Bon24}. }\label{beta}
\end{figure}

\begin{figure}[h]
\centering
\includegraphics[width=0.66\textwidth]{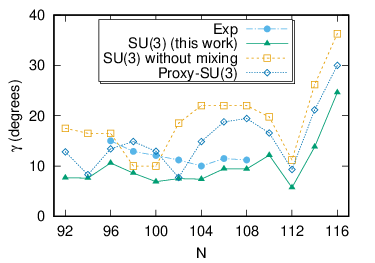}
\caption{Pseudo-SU(3) model predictions (labelled “SU(3) (this work)”) for the collective deformation variable $\gamma$ along the $^{164-188}$Hf isotopic chain, compared with the empirical values (“Exp”) \cite{Bon17}, the results obtained considering only the most symmetric pseudo-SU(3) representation (“SU(3) without mixing”) \cite{Bon20}, and the results from the proxy-SU(3) scheme \cite{Bon24}. }\label{gamma}
\end{figure}

Using the basis composed of the states with the highest spatial symmetry for $\tilde{S}_\alpha = 0$ and 1 (with $\alpha = \pi, \nu$) and the Hamiltonian discussed in Sec.~\ref{sec-ham}, it is possible to calculate various properties of the lowest-energy states in $^{164-188}$Hf. Figures~\ref{beta} and \ref{gamma}
show the results for the geometric variables $\beta$ and $\gamma$, respectively.

The predictions obtained for the quadrupole deformation $\beta$ (Fig.~\ref{beta}) in the present work (solid triangles) are compared with the experimental data (solid circles), as well as with the results from the pseudo-SU(3) model taken as a dynamical symmetry (open squares, using only the irrep with highest spatial symmetry) \cite{Bon20} and those from the proxy-SU(3) approach (open diamonds) \cite{Bon24}. Only the most symmetric irreps (labeled hw) of Ref. \cite{Bon24} were used in the comparison, not the next-to-most-symmetric ones. The most symmetric irrep dominates the collective properties, while the next-to-most-symmetric irrep becomes relevant only when the most symmetric irrep has $\mu=0$, which is not the case for the nuclei considered here. 

A common trend in $\beta$ is observed in the experimental data and in the three calculations, reproducing the experimental maximum of the deformation at $N=100$, with a decrease as one moves away from mid-shell. However, it should be noted that the $\beta$ values obtained in the present contribution are approximately 20\% smaller than the experimental data in the central region of the shell. This discrepancy could be due to the absence of nucleons in the intruder sector.

The experimental value $\beta = 0.182$ at $N=92$ stands out, as it lies well below all theoretical predictions. For neutron numbers close to the prolate–oblate transition ($N \sim 114, 116$), no experimental data are available, but the models agree on numerical values close to $\beta = 0.200$.

Figure~\ref{gamma} shows the results for the variable $\gamma$ from the present model, together with the empirical values and the calculations with the dynamical symmetry and proxy-SU(3) schemes. For isotopes in the region from $N=92$ to $N=110$, nearly constant values are obtained with the present model (around $8^\circ$–$12^\circ$), lying very close to the experimental values and showing only small variations. In this region, the dynamical-symmetry and proxy-SU(3) calculations predict values that fluctuate strongly within $7^\circ < \gamma < 22^\circ$. For $N \ge 112$, where no experimental data are available, the model predicts a decrease to $\gamma = 5.71^\circ$ at $N=112$, followed by a significant increase to $\gamma = 13.8^\circ$ and $24.6^\circ$ at $N=114$ and 116, respectively. Similar variations in $\gamma$ are also predicted by the dynamical-symmetry and proxy-SU(3) approaches. Thus, although the ground state is somewhat triaxial throughout the region, with typical values of $\gamma \sim 10^\circ$, a clearly triaxial deformation is predicted around $N \sim 116$.

The $\gamma$ values predicted in the present work differ qualitatively from those in Refs. \cite{Bon20,Bon24}. Those earlier works relied on the most symmetric irrep alone (or a simple average of two irreps) without diagonalizing a realistic Hamiltonian, leading to pronounced oscillations in $\gamma$ as a function of neutron number. In contrast, the present work considers an extended Hilbert space that includes mixing among several pseudo-SU(3) irreps, driven by the Nilsson potential and the pairing interaction. Diagonalizing a realistic Hamiltonian with these symmetry-breaking terms smooths out the sharp oscillations, yielding more stable $\gamma$ values that are closer to empirical systematics.

\section{Effect of Nilsson and pairing terms on the collective deformation variables}
\label{sec-pair}

We also aim to clarify how the interaction terms favour the transition from a prolate to a triaxial shape at \(N=116\), and how the values of \(\gamma\) change as the shell is filled. In this Section, we present the values of \(\beta\) and \(\gamma\) for the same isotope chain, but now with selected interaction terms in the Hamiltonian of Eq.(\ref{eq:ham}) switched off. As discussed in previous works \cite{Var98,Var-00,Hir99,Zuk95}, the low-energy properties are governed by the interplay between the Nilsson single-particle and quadrupole–quadrupole terms, which determine the main components of the wave function and thus all observables.

\begin{figure}[h]
\centering
\includegraphics[width=0.76\textwidth]{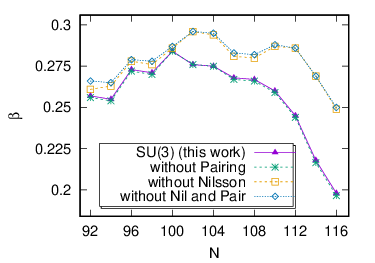}
\caption{Predictions of the pseudo-SU(3) model for the variable $\beta$ as a function of the Hamiltonian used. The solid triangles correspond to the same values shown in Fig.~\ref{beta}. The other three curves display the values of $\beta$ when the indicated terms are removed from the Hamiltonian.}\label{be-ham}
\end{figure}

\begin{figure}[h]
\centering
\includegraphics[width=0.76\textwidth]{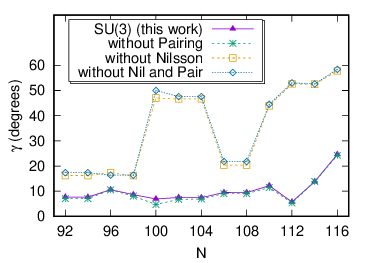}
\caption{Predictions of the pseudo-SU(3) model for the variable $\gamma$ as a function of the Hamiltonian used. The solid triangles correspond to the same values shown in Fig.~\ref{gamma}. The other three curves display the values of $\gamma$ when the indicated terms are removed from the Hamiltonian.}\label{ga-ham}
\end{figure}

To address this point, Figs. \ref{be-ham} and \ref{ga-ham}
show the calculated values of the geometric variables \(\beta\) and \(\gamma\), respectively, as a function of the Hamiltonian used. In both plots, the solid line with filled triangles labelled “SU(3) (this work)” displays the results of the present model with the full Hamiltonian discussed in Sec.\ref{sec-ham}. The dotted line with stars, labelled “without Pairing”, shows the values of \(\beta\) and \(\gamma\) obtained by setting \(G_\alpha = 0\) (with \(\alpha = \pi, \nu\)) in Eq.(\ref{eq:ham}), i.e., excluding pairing correlations.  Hence, we conclude that the calculated geometric variables are essentially insensitive to pairing in this case, as these two curves are nearly identical. The curves with open squares give the values obtained without the Nilsson single-particle terms (\(C_\alpha = D_\alpha = 0\) in Eq.(\ref{eq-hsp})), while the curve with open diamonds corresponds to the Hamiltonian without either the Nilsson or pairing terms. 

As can be seen in both Figures, the absence of the Nilsson single-particle terms has a major impact on the values of $ \beta$ and $\gamma$, confirming what has been pointed out in previous works \cite{Var98,Var-00}, namely that the interplay between the single-particle and quadrupole–quadrupole terms in the Hamiltonian governs the SU(3) mixing in the wave functions. The delicate balance between these two interactions, whose strengths are taken from known systematics and not used as fitting parameters, determines the values of the $\beta$ and $\gamma$ variables in the $^{164-188}$Hf nuclei. On the other hand, the pairing interaction strongly mixes the SU(3) irreps \cite{Tro95,Bah95,Esc98} and plays an important role in determining the moments of inertia of deformed nuclei, whereas its effect on the wave function and on the values of the geometric variables $\beta$ and $\gamma$ is marginal.

This result clarifies the point raised in Ref.~\cite{Bon17} regarding the oscillations in the value of $\gamma$ in the isotopes of this region. On page 8 of that work, the authors state that “It is our speculation that this (rather strong oscillation) reflects the effects of pairing which are ignored here and which will tend to spread the orbit occupancies in the ground state ... and mute these oscillations”. The results presented here (Fig.~\ref{gamma}) allow us to conclude that the oscillations in $\gamma$ are not due to the absence of pairing, but rather to the omission of the Nilsson single-particle terms.

\section{Summary}\label{sec-conclu}

Using the pseudo-SU(3) nuclear model, we describe the shape observables $\beta$ and $\gamma$ along the $^{164-188}$Hf isotope chain. The model space includes the states with maximal spatial symmetry, $(\lambda+\mu)_{\text{max}}$ and $r(\lambda,\mu)_{\text{max}}$, and is further enriched by adding other states with pseudo-spin 0 and 1 whose irreps satisfy $\lambda' + \mu' = (\lambda + \mu)_{\text{max}} - 1, (\lambda + \mu)_{\text{max}} - 2, \ldots$, and which possess the largest possible symmetry fraction $r$. The Hamiltonian contains symmetry-breaking terms, parametrised with a normalisation dictated by the universal $A^{-1/3}$ scaling, and involves no free parameters. The results indicate that the model predicts a clear dominance of prolate deformation, with a moderate degree of triaxiality ($8$–$12^\circ$) in Hf nuclei at the beginning and in the middle of the shell, and a pronounced triaxiality ($\gamma = 24.6^\circ$) in nuclei approaching the shell closure. Furthermore, the maximum in the $\beta$ deformation at $N = 100$ is correctly reproduced, although with a value about 20\% smaller than the experimental one, possibly as a consequence of the absence in the model of nucleons in intruder-parity orbitals. This constitutes the main limitation of the present approach.

The results clearly show the importance of considering a Hamiltonian that includes the Nilsson, quadrupole–quadrupole, and pairing terms. It is demonstrated that the interplay among these three terms produces predominantly prolate states over almost the entire Hf chain, with a clear dominance of triaxial shapes in nuclei with $N \ge 112$. 

The analysis of the Nilsson and pairing terms in the Hamiltonian on the deformation parameters $\beta$ and $\gamma$ shows that the Nilsson energies play a predominant role in determining the nuclear structure and the emergence of triaxiality, whereas the pairing term has only a marginal effect. 

The fundamental ingredients of the model are: (i) the consideration of states with high spatial symmetry and the highest possible symmetrisation $r$ in the subspace spanned by pseudo-spin 0 and 1 states; (ii) a realistic Hamiltonian composed of the most relevant terms, parametrised according to the systematic; and (iii) a mixing of the more symmetric irreducible representations generated by the breaking of pseudo-spin symmetry, mainly through the Nilsson single-particle energies. This model can be applied to other rare-earth isotope chains, which is left for future work.

No funding was received for conducting this study. The authors have no competing interests which are relevant to the content of this article to declare.

\end{document}